\documentclass[twocolumn,superscriptaddress,showpacs,aps,amsmath,amssymb,prl]{revtex4}
\usepackage{amsfonts}
\usepackage{amsmath}
\usepackage{amssymb}
\usepackage{bm}
\usepackage{graphicx}

\newcommand{\nl}{\nonumber \\}
\newcommand{\be}{\begin{equation}}
\newcommand{\ee}{\end{equation}}
\newcommand{\bea}{\begin{eqnarray}}
\newcommand{\eea}{\end{eqnarray}}
\newcommand{\Fig}[1]{Fig.\,\ref{#1}}
\newcommand{\Eq}[1]{Eq.\,(\ref{#1})}
\newcommand{\Eqs}[1]{Eqs.\,(\ref{#1})}

\newcommand{\br}{\mathbf{r}}
\newcommand{\ep}{\epsilon}

\begin{document}

\title{Time-dependent density-functional theory for real-time electronic dynamics on material surfaces}

\author{Rulin Wang}
\affiliation{Hefei National Laboratory for Physical Sciences at the
 Microscale, University of Science and Technology of China, Hefei,
 Anhui 230026, China}
\affiliation{Department of Physics, University of Science and
Technology of China, Hefei, Anhui 230026, China}

\author{Dong Hou}
\affiliation{Hefei National Laboratory for Physical Sciences at the
 Microscale, University of Science and Technology of China, Hefei,
 Anhui 230026, China}

\author{Xiao Zheng} \email{xz58@ustc.edu.cn}
\affiliation{Hefei National Laboratory for Physical Sciences at the
Microscale, University of Science and Technology of China, Hefei, Anhui
230026, China}


\date{July~22, 2013}


\begin{abstract}

The real-time electronic dynamics on material surfaces is critically important to a variety of applications. However, their simulations have remained challenging for conventional methods such as the time-dependent density-functional theory (TDDFT) for isolated and periodic systems.
By extending the applicability of TDDFT to systems with open boundaries, we achieve accurate atomistic simulations of real-time electronic response to local perturbations on material surfaces.
Two prototypical scenarios are exemplified: the relaxation of an excess electron on graphene surface, and the electron transfer across the molecule-graphene interface. Both the transient and long-time asymptotic dynamics are validated, which accentuates the fundamental importance and unique usefulness of an open-system TDDFT approach. The simulations also provide insights into the characteristic features of temporal electron evolution and dissipation on surfaces of bulk materials.

\end{abstract}

\maketitle
%

How electrons evolve at the surfaces or interfaces of materials is fundamentally significant to a variety of applications, including photovoltaics, nanoelectronics, heterogeneous catalysis, etc.
Consider a prototypical system that a molecule is adsorbed on a surface of a material.
For instance, in a dye-sensitized solar cell \cite{Ore91737}, photo-excited electrons transfer from the dye molecule to the semiconductor surface, and then drain into the bulk \cite{Hag00269}.
In a biomimetic water-splitting complex, a catalytic molecule acquires electrons by oxidation of water and then feeds them into the supporting conductor \cite{Li09230}. Apparently, for these systems the real-time electronic processes on material surfaces are crucially important to their functionality.
Accurate simulations at atomic level will be very helpful for understanding the key features of the real-time electronic dynamics and the underlying mechanisms.

Considering the size and complexity of a system involving material surface, the time-dependent density-functional theory (TDDFT) \cite{Run84997,Mar04427,Bur05062206,Cas12287} is potentially suitable for carrying out the theoretical studies, due to its favorable balance between accuracy and efficiency.

For any practical application of TDDFT, a boundary condition exists explicitly or implicitly. So far the success of TDDFT has been largely restricted to isolated and periodic boundary conditions.
For isolated systems (atoms, molecules, clusters, etc.) the electron density falls off to zero at infinite distance, while for periodic systems (polymers, crystals, etc.) the electron density possesses the lattice translational invariance symmetry.
It is worth pointing out that, although TDDFT has been employed to study excited-state properties (such as absorption and electron-energy-loss spectra) of periodic solids \cite{Wai0010149,Ole015962,Oni02601,Bot07357}, a rigorous proof justifying the existence of a TDDFT for periodic systems is still
lacking \cite{Bae08044103,Li08044105,Ver08166401}.

Apparently, for a composite system where a molecule is adsorbed on a material surface, neither an isolated nor a periodic model is suitable, particularly when the electronic dynamics is triggered by a local perturbation. In such a case, it is ideal to treat the molecule with the part of surface around the adsorption site as an \emph{open} system, while taking the rest of bulk material as the \emph{environment}.

TDDFT for open systems has been proposed to study electron transport through molecular or nanoelectronic devices coupled to macroscopic electrodes
\cite{Kos031,Kur05035308,Bur05146803,Zhe07195127,Li07075114}.
With electron current flowing through a device, it is impractical or inappropriate to treat the device-electrodes composite or the device itself as either isolated or periodic.
With the open-system TDDFT, the device is regarded as a system having an \emph{open} boundary, while the electrodes constitute the \emph{environment} which serves as electron reservoirs and energy sinks.
The open-system TDDFT can be built on a formally exact theoretical foundation
\cite{Ste0414,Ven048025,Zhe05a,Zhe07195127,Zhe1114358,Yue10043001,Tem11074116}.
In particular, the existence of a rigorous TDDFT for a general open system coupled to any large but finite environment has been proved by a time-dependent holographic electron density theorem \cite{Zhe05a,Zhe07195127,Zhe1114358}.
In principle (but unfortunately not in practice), the electron density inside the open boundary alone should suffice to determine all equilibrium and nonequilibrium properties of the entire composite system (system plus
environment) \cite{Zhe1114358}.
The main challenge for the open-system TDDFT is the accurate characterization of dissipative processes occurring at the designated boundary, including the energy relaxation, electron transfer, and decoherence.
So far the applications have been limited to electron transport in one-dimensional systems
\cite{Ste04195318,Yam08495203,Ste10115446,Xin10205112,Ke10234105,Wen115519},
because it is difficult to resolve the atomistic and spectral details of environment of higher dimensions. Meanwhile, addressing their influences on the open system presents further challenges.

In this letter, we will (1) show how the effects of bulk surface (environment) can be taken into account accurately and efficiently in TDDFT; (2) extend the applicability of TDDFT to simulations of real-time electronic dynamics on two-dimensional material surfaces; (3) elucidate the boundary effects on the simulation results; and (4) highlight the fundamental importance and uniqueness of the open-system model and approach.

For isolated systems, the Kohn--Sham equation of motion for the reduced single-electron density matrix is \cite{Yam03153105},
\be
    i \dot{\bm \sigma}(t) = [ \bm h(t), \bm \sigma(t) ]. \label{tdks_iso}
\ee
Here, $\bm \sigma(t)$ and $\bm h(t)$ are the Kohn--Sham density matrix and Fock matrix of the isolated system, respectively.
While for periodic systems we have \cite{Cas062465}
\be
   i \dot{\bm \sigma}_{\bm k}(t) = [ \bm h_{\bm k}(t), \bm \sigma_{\bm k}(t) ], \label{tdks_per}
\ee
where $\bm k$ is the wavevector, and $\bm\sigma(t) = \int_{\scriptscriptstyle \rm BZ} \bm\sigma_{\bm k}(t)\, d\bm k / \Omega_{\scriptscriptstyle \rm BZ}$ with $\Omega_{\scriptscriptstyle \rm BZ}$ being the volume of Brillouin zone.

For open systems, the general Kohn--Sham equation of motion has been derived as \cite{Zhe07195127}
\be
  i \dot{\bm \sigma}(t) = [\bm h(t), \bm \sigma(t)] - i \bm Q(t).
  \label{tdks_open}
\ee
Here, $\bm \sigma(t)$ and $\bm h(t)$ are of the size of the open system, and the matrix $\bm Q(t)$ addresses the effects of environment on the system inside the boundary.
The main challenge is to find an accurate and efficient scheme to evaluate $\bm Q(t)$.
Various approaches have been proposed, such as the nonequilibrium Green's function formalism \cite{Kur05035308,Zhu05075317}, the adiabatic wide-band limit approximation \cite{Zhe07195127}, and the perturbative master equation approach \cite{Li07075114,Cui06449}. Recently, a hierarchical equations of motion (HEOM) approach based on a formally exact quantum dissipation theory \cite{Jin08234703,Zhe121129} has been developed \cite{Cro09245311,Zhe10114101,Xie12044113,Tia12204114,Zha13085110}.

In the HEOM scheme, if the environment consists of only one electron reservoir, $\bm Q(t)$ is resolved by
\be
     \bm Q(t) = -i \sum_{m=1}^{M} \left[\bm \varphi_{m}(t) - \bm \varphi^\dag_{m}(t) \right].   \label{q_heom}
\ee
Here, $\{\bm \varphi_{m}(t)\}$ are the first-tier auxiliary density matrices, and their definitions are not unique \cite{Zhe10114101}. To resolve $\bm Q(t)$ accurately with a minimal number ($M$) of matrices, we adopt a Pad\`{e} decomposition algorithm for the Fermi function \cite{Hu10101106,Hu11244106} and a multi-Lorentzian expansion \cite{Zhe10114101,Xie12044113} for the reservoir spectral functions \cite{Sup1}.
$\{\bm \varphi_{m}(t)\}$ satisfy the equations of motion as follows,
\begin{align}
  i\,\dot{\bm{\varphi}}_m(t) &= \left[ \bm h(t) - \Delta\mu(t) - i\gamma_{m}^{+} \right] \,\bm \varphi_{m} + \sum_{m'=1}^{M} \bm \psi_{mm'} \nl
  &\quad -i \left[ (1 - \bm\sigma) \bm A^{<+}_m + \bm\sigma \bm A^{>+}_m \right], \label{phi_heom} \\
  i\,\dot{\bm{\psi}}_{mm'}(t) &= i \left( \gamma^{-}_{m'} - \gamma^{+}_m \right) \bm \psi_{mm'} + i \left( \bm A^{>-}_{m'} - \bm A^{<-}_{m'} \right) \bm \varphi_m \nl
  &\quad -i \bm \varphi^\dag_{m'} \left( \bm A^{>+}_m - \bm A^{<+}_m \right). \label{psi_heom}
\end{align}
Here, $\{\bm \psi_{mm'}(t)\}$ are the second-tier auxiliary density matrices,
$\Delta \mu(t)$ is the variation of reservoir chemical potential, and the matrices $\{\bm A^{<\pm}_m, \bm A^{>\pm}_m \}$ and coefficients $\{\gamma_m^\pm\}$ are obtained from reservoir spectral decomposition \cite{Sup1}.
Obviously, \Eqs{tdks_open}--\eqref{psi_heom} form a closed set of equations for the unknowns $\{\bm \sigma, \bm \varphi_m, \bm \psi_{mm'}\}$, which determine completely the real-time dynamics of the open system.

\begin{figure}
\includegraphics[width=\columnwidth]{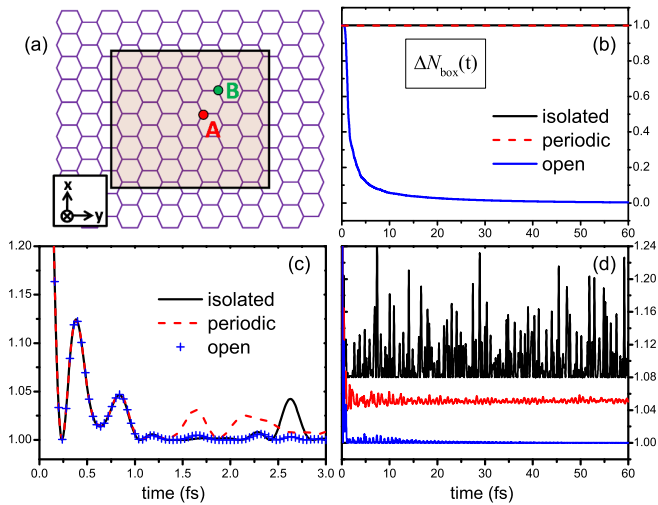}
\caption{(Color online). (a) Schematic diagram of a graphene surface. The shaded area represents the simulation box which is treated explicitly by TDDFT approaches subjected to specific boundary conditions. (b) Number of excess electron that remains within the box versus time upon its injection onto atom A at $t = 0$. (c) and (d) display the variation of electron occupation on atom A ($n_{\rm A}$) from $t = 0$ to $3\,$fs and to $60\,$fs, respectively. For clarity, in (d) the lines of isolated and periodic models are elevated by 0.08 and 0.04, respectively.
}
\label{fig1}
\end{figure}

We first investigate the real-time electronic response of a graphene monolayer
to a local perturbation. As shown in \Fig{fig1}(a), the bulk graphene resides in the \emph{xy} plane. A rectangular piece of graphene containing 96 atoms is chosen as the simulation box (the shaded area), which is treated explicitly by TDDFT approaches subjected to specific boundary conditions.
The atoms of the graphene form a perfectly periodic two-dimensional lattice, and the C--C bond length assumes the typical value of $1.42\,$\AA\,\cite{Bou08035427}. Because of the delocalized \emph{sp}$^2$ network, conduction electrons can move easily on the (\emph{xy}) plane \cite{Saj13a}.
A tight-binding Hamiltonian is used to represent the $p_z$ electrons, with the on-site energy $\ep_0 = 0$ and nearest-neighbor coupling $\gamma = 2.7\,$eV \cite{Net09109}.

To apply the HEOM approach to simulate the real-time electronic dynamics on graphene surface, the spectral function of bulk graphene outside the simulation box (environment) needs to be evaluated accurately. This is achieved by using a reciprocal-space sampling technique of Ref.\,[\onlinecite{Zhe10114703}]. Moreover, to ensure high accuracy is achieved for the resulted real-time dynamics, a residual correction technique is adopted to compensate any minor error resulted from the decomposition of \Eq{q_heom} \cite{Sup1}.

The graphene is initially in its equilibrium state. At a certain time (set as $t = 0$), an excess electron is injected onto atom A inside the simulation box; see \Fig{fig1}(a). Such a local perturbation drives the graphene out of equilibrium, and induces electronic response that propagates outward from the perturbed site.
Experimentally the single-electron injection may be realized with a tunneling junction setup \cite{Fev071169}.
The present scenario can be deemed as a quantum analog of ``dripping a droplet into a liquid'', which results in generation and spreading of electron ripples on the graphene surface.

Dissipation of an excess electron on graphene plane is supposed to be one of the simplest examples of electronic dynamics on material surfaces.
However, even such a simple case turns out to be rather challenging for conventional TDDFT. The major difficulty is that in reality the number of excess electron remaining in the box, $\Delta N_{\rm box}(t) = \sum_{a\in {\rm box}} \bm \sigma_{aa}(t) - \bm \sigma_{aa}^{\rm eq}$, is not conserved as time goes on; whereas the isolated or periodic boundary condition forces it to be a constant; see \Fig{fig1}(b).
In contrast, the open-system TDDFT allows the excess electron to propagate through the (artificially designated) boundary, and properly dissipate into the surrounding bulk graphene. A characteristic dissipation time $\tau_d$ can be defined as the time that $\Delta N_{\rm box}$ reduces to less than $0.01$. From \Fig{fig1}(b), we have $\tau_d = 37.9\,$fs.

The simulation proceeds as follows. At $t = 0$ the electron occupation on atom A ($n_{\rm A} = \bm \sigma_{\rm AA}$) increases instantaneously by $1$, \emph{i.e.}, $n_{\rm A}(0^+) = n_{\rm A}^{\rm eq} + 1$. The subsequent real-time electronic dynamics are obtained by solving \Eqs{tdks_iso}--\eqref{tdks_open} for isolated, periodic, and open boundary conditions, respectively. For simplicity the system is treated as spin-closed at any time.

We now examine how the electron occupation on atom A varies in time.
The short-time evolution of $n_{\rm A}(t)$ is shown in \Fig{fig1}(c). At $t < 1\,$fs, the lines associated with the three types of boundary conditions overlap perfectly with each other. Since the electronic dynamics should be identical before the first ripple wavefront reaches the boundary, such quantitative agreement verifies the accuracy of our proposed HEOM approach. Approximate schemes such as the wide-band limit \cite{Zhe07195127} and complex absorbing potential \cite{Bae043387,Var11195130,Zha13205401} have been used for simplifying the evaluation of $\bm Q(t)$, which should also be tested carefully to ensure the correct short-time transient dynamics be reproduced.

Regarding the electronic response of graphene, a fundamental quantity is the time scale that the ripples dissipate away so that the equilibrium is restored. To this end, the long-time electronic dynamic are displayed in \Fig{fig1}(d). In the isolated model the electron density persists a significant and persistent fluctuation due to the fully reflective boundary. In the periodic model the fluctuation retains a residual yet nonvanishing amplitude. This is because a ripple leaving from one side of the box is forced to reenter at the counter side, so that the excess electron is ``evened'' out in the box. Apparently, it is only with the open boundary that the fluctuation eventually damps out. Similar phenomena are observed if the local perturbation is instead an abrupt shift of an atomic potential energy; see Ref.\,[\onlinecite{Sup1}].

If one insists on using conventional TDDFT for isolated or periodic systems to study the real-time electronic response to a local perturbation, the size of the box must be enlarged drastically to avoid the unwanted boundary effects. In practice, such calculations could be exceedingly costly, if not impossible.

\begin{figure}
\includegraphics[width=\columnwidth]{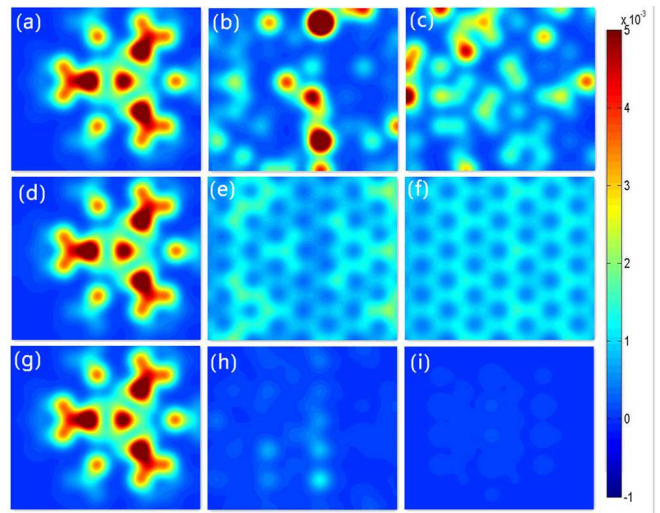}
\caption{(Color online). Snapshots of time-dependent density of excess electron inside the box, $\Delta n(\br, t) = n(\br, t) - n^{\rm eq}(\br)$, after an electron is injected onto the GML at $t = 0$. The upper, middle, and lower rows correspond to isolated, periodic, and open systems; and the left, middle, and right columns are for time instants $t = 0.5$, $5$, and $20\,$fs, respectively.
}
\label{fig2}
\end{figure}

The evolution of \emph{excess} electron density, $\Delta n(\br, t) = n(\br, t) - n^{\rm eq}(\br)$, is visualized in \Fig{fig2}, with local $p_z$ orbitals represented by Gaussian functions. At $t = 0.5\,$fs when the propagation has not reached the boundary of box, all three models give the same pattern of ripples. The $\Delta n(\br, t)$ exhibits a distinct three-fold symmetry on the \emph{xy} plane, reflecting the $sp^{2}$ bonding characteristics. At $t = 5\,$fs, a significant portion of excess electron should have dissipated into the surrounding graphene. While the open system model allows the excess electron to permeate through the boundary, both isolated and periodic models constrain the electron from leaving the box. At $t$ as long as $20\,$fs, the ripples of the isolated model remain fluctuating; while the periodic boundary makes the excess electron ``evenly'' distributed among the $sp^2$ network. Only with the open system model can the excess electron dissipate away properly.

We now move to a more complex scenario where a four-atom linear molecule is adsorbed on the graphene surface at atom B. TDDFT has been employed to study the ultrafast electron transfer in dye-sensitized solar cells with the isolated \cite{Guo0816655} and periodic \cite{Dun078528,Men083266} models. However, to the best of our knowledge, the boundary effects on the resulting real-time dynamics have not been explored.

\begin{figure}
\includegraphics[width=\columnwidth]{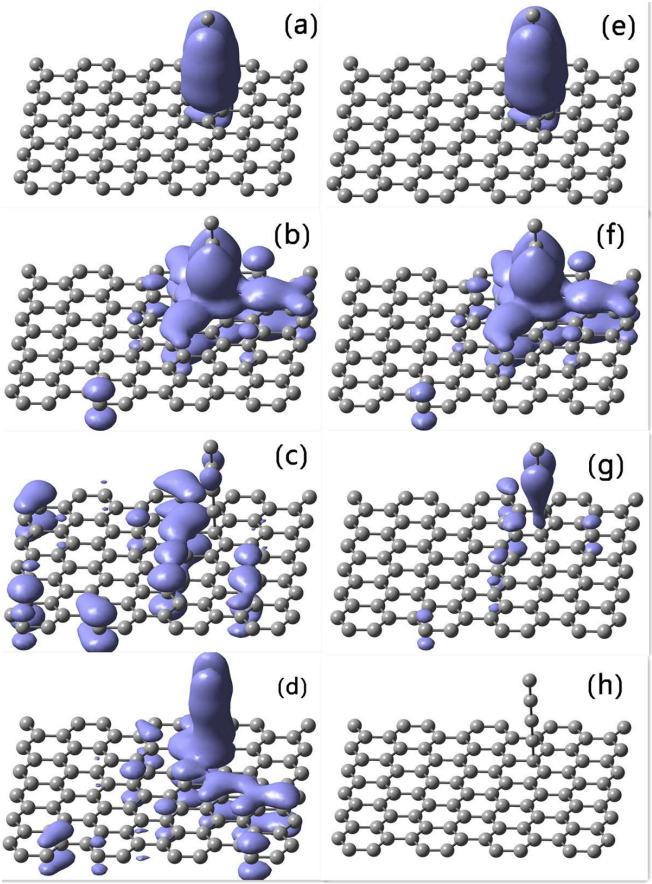}
\caption{(Color online). Isosurfaces of excess electron density $\Delta n(\br,t) = 4\times10^{-4}\,{\rm bohr}^{-3}$ for isolated (left column) and open (right column) system models at various time instants and $\lambda = \gamma = 2.7\,$eV. The four rows from top to bottom are snapshots at $t = 0.5$, $1$, $2.5$, and $6.5\,$fs, respectively.
}
\label{fig3}
\end{figure}

The molecule assumes the same values of $\epsilon_0$ and $\gamma$ as the graphene, and their bonding strength is $\lambda$.
At $t = 0$ an excess electron enters at the free end of molecule, which drives the molecule-graphene composite out of equilibrium. Again, \Eqs{tdks_iso}--\eqref{tdks_open} are used to simulate the real-time electronic dynamics subjected to the isolated, periodic, and open boundary conditions, respectively.

The evolution of excess electron is displayed in \Fig{fig3} for $\lambda = \gamma$, where snapshots of isolated boundary at various time instants are compared with the open-system counterparts. At $t < 1\,$fs (upper two rows), the dynamics occurs mostly inside the box, and both models give essentially the same result. In contrast, at $t > 2.5\,$fs (lower two rows) the dynamics become qualitatively different: in the isolated model the electron gets reflected at the boundary and sometimes repopulates on the molecule; while with the open boundary the excess electron gradually drains into the surrounding bulk graphene and vanishes entirely in the long-time limit.

\begin{figure}
\includegraphics[width=\columnwidth]{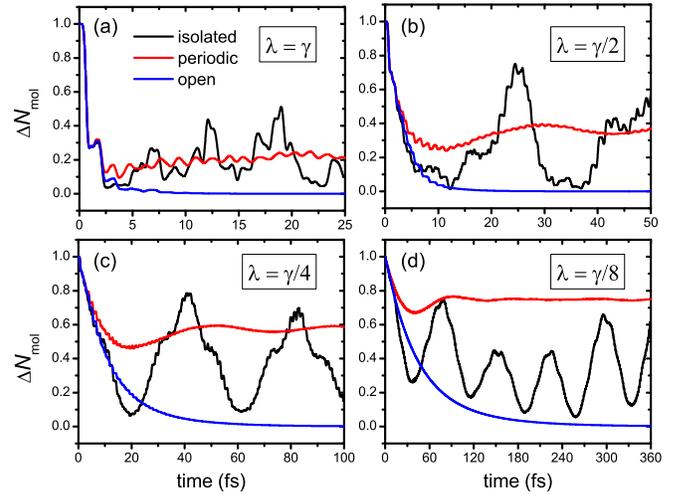}
\caption{(Color online). Number of excess electrons on the molecule versus time, $\Delta N_{\rm mol}(t)$, for the three boundary conditions and at (a) $\lambda = \gamma$, (b) $\lambda = \gamma/2$, (c) $\lambda = \gamma/4$, and (d) $\lambda = \gamma/8$, respectively.
}
\label{fig4}
\end{figure}

Figure~\ref{fig4} plots the population of excess electron on the molecule versus time, $\Delta N_{\rm mol}(t) = \sum_{a\in {\rm mol}} \bm\sigma_{aa}(t) - \bm\sigma^{\rm eq}_{aa}$, at various values of $\lambda$. Apparently, only with the open system model the electron relaxation exhibits the correct long-time asymptotic behavior.
A characteristic transfer time $\tau_t$ can be defined as the time that $\Delta N_{\rm mol}$ reduces to less than $0.01$. From \Fig{fig4}, $\tau_t$ is evaluated to be 7.6, 15.8, 66.5, and 270.5\,fs for $\lambda/\gamma = 1$, $1/2$, $1/4$, and $1/8$, respectively. This clearly infers that $\tau\propto 1/\lambda^2$, except at $\lambda = \gamma$ where the intramolecular dynamics is nearly in resonance with the interfacial electron transfer modes.
While the isolated model gives reasonable short-time transient dynamics, at a long time the excess electron reappears on the molecule, which is clearly due to the artificial quantum confinement effect.
In contrast, the periodic boundary withholds the electron from leaving the molecule, and even more so with a smaller $\lambda$. This stresses the fact that, the periodic (or isolated) model lacks a relaxation channel in the absence of external fields and dissipative baths. Consequently, the excess electron is unable to lose energy to populate onto the low-lying graphene states.

To conclude, we have proposed an HEOM approach to achieve accurate atomistic simulation of real-time electronic dynamics on two-dimensional graphene surface, which greatly extends the practicality of TDDFT. The nontrivial boundary effects accentuate the necessity and advantages of an open-system approach. The simulations shed important lights on the characteristic features of prototypical real-time electronic dynamics on material surfaces. Certainly, more work is needed to go beyond a model Hamiltonian, and to meet other challenges such as the inclusion of excitonic and plasmonic effects.

\acknowledgments

The support from the NSF of China (No.\,21103157 and No.\,21233007), the Fundamental Research Funds for Central
Universities (No.\,2340000034 and No.\,2340000025), and the Strategic Priority Research Program (B) of the CAS (XDB01020000) is gratefully
acknowledged.
We thank Prof. GuanHua Chen and Prof. YiJing Yan for stimulating discussions.


%
%

\end{document}